**Distinct Surface and Bulk Superconductivity in the Kagome Superconductor $SrSn_3$**

Qun Zhu[1], Yong-Wei Wang[1], Ji-Hai Zhang[1], Qiang-Jun Cheng[1], Chen-Yu Hu[2], Jun-Zhong Wang[2],

Qi-Kun Xue[1,3,4,5,6], Xu-Cun Ma[1,3,*], Can-Li Song[1,3,*]

[1]*Department of Physics and State Key Laboratory of Low-Dimensional Quantum Physics, Tsinghua University, Beijing 100084, China*

[2]*School of Physical Science and Technology, Southwest University, Chongqing 400715, China*

[3]*Frontier Science Center for Quantum Information, Beijing 100084, China*

[4]*Shenzhen Institute for Quantum Science and Engineering and Department of Physics, Southern University of Science and Technology, Shenzhen 518055, China*

[5]*Beijing Academy of Quantum Information Sciences, Beijing 100193, China*

[6]*Hefei National Laboratory, Hefei 230088, China*

Surface and bulk superconductivity may possess fundamentally different superconducting properties in quantum materials with nontrivial electronic structures, yet their superimposed spectroscopic signatures often prevent direct experimental access to each superconducting channel. Here we reveal, in epitaxial films of the kagome superconductor $SrSn_3$, distinct surface and bulk superconducting channels with markedly different superconducting gaps, upper critical fields, and vortex-core electronic states by tuning the tunneling junction resistance in scanning tunneling spectroscopy. The surface superconductivity is characterized by a thickness-independent superconducting gap and an enhanced upper critical field, whereas the bulk superconducting channel exhibits a larger superconducting gap that decreases with reducing film thickness and a much lower upper critical field. Within magnetic vortex cores, robust non-split zero-bias conductance peaks are observed exclusively in the surface superconducting channel, while pronounced zero-bias suppression is consistently associated with the bulk superconducting channel. These findings demonstrate that the vortex-core electronic structure depends sensitively on the underlying superconducting channel, providing new insight into vortex-bound states in topological quantum materials.

[*]*Corresponding authors. Emails: clsong07@mail.tsinghua.edu.cn,* xucunma@mail.tsinghua.edu.cn

Quantum materials with complex electronic structures often host multiple electronic channels associated with surfaces, interfaces, and the bulk [1-3]. When superconductivity develops simultaneously in surface (or interface) and bulk channels, the resulting superconducting states can exhibit distinct dimensionalities, pairing strengths, and magnetic responses. Such a surface-bulk superconducting dichotomy has been suggested in a variety of systems, including potential topological superconductors (TSCs) [1,4-9], Weyl semimetals [10], and kagome superconductors [11,12]. Among these compounds, intrinsic and engineered TSCs have attracted particular attention because they provide a promising platform for realizing Majorana zero modes for fault-tolerant quantum computation [13-20]. Experimentally, several TSC candidates and Weyl semimetals have been reported to exhibit surface superconducting gaps ($\Delta_s$) considerably larger than their bulk counterparts ($\Delta_b$) [8,15, 21-24], frequently accompanied by enhanced upper critical fields ($B_{c2}$) [24]. Despite extensive experimental and theoretical efforts, however, establishing a direct correspondence between the underlying superconducting channels and their characteristic spectroscopic signatures remains an outstanding challenge. More recently, kagome superconductors have emerged as an attractive platform for investigating the interplay between superconductivity, nontrivial topology, and electron correlations [25-27]. While significant progress has been achieved in the $AV_3Sb_3$ ($A$ = Cs, Rb, K) family [28], including signatures of vortex-bound zero-bias conductance peaks (ZBCPs) and possible topological superconductivity [27,29,30], whether distinct surface and bulk superconducting channels exist in kagome superconductors, and if so, how they are connected to their respective superconducting and vortex properties, remain largely unexplored.

Scanning tunneling spectroscopy (STS) provides direct access to the electronic density of states (DOSs) with atomic-scale spatial resolution and has become one of the most powerful probes of superconductivity [31]. In quantum materials hosting coexisting surface and bulk superconductivity, however, the measured differential conductance (d$I$/d$V$) spectrum generally reflects the combined DOSs from both superconducting channels [15,21-24]. Consequently, key quantities such as the superconducting gap, upper critical field ($B_{c2}$), and vortex-core electronic structure remain difficult to assign unambiguously to the distinct surface and bulk superconducting channels. Here, we overcome this challenge by systematically tuning the tunneling junction resistance ($R_N$), or equivalently the tip-sample distance ($d$), in high-resolution cryogenic STS measurements (see Supplemental Material, Fig. S1 [32]). This enables continuous spectroscopic separation of the surface and bulk superconducting channels in the kagome superconductor $SrSn_3$, uncovering their fundamentally different superconducting gaps, upper critical fields, and vortex-core electronic structures.

The experiments were performed using a Unisoku ultrahigh-vacuum (UHV) low-temperature scanning tunneling microscopy (STM, USM1300) integrated with a molecular beam epitaxy (MBE) system for *in-situ* sample preparation. Commercial Si(111) wafers with a resistivity of < 0.005 Ω·cm were used as substrates. Prior to film preparation, the substrates were degassed at 600°C overnight and subsequently flash-annealed to 1200°C approximately 20 times to obtain a well-defined Si(111)-(7 × 7) reconstructed surface. High-purity Sr (99.95%) and Sn (99.9999%) were co-deposited from standard Knudsen effusion cells on the Si(111)-(7 × 7) substrates maintained at 200°C, followed by post-growth annealing at ~ 500°C. The resulting high-quality $SrSn_3$ films were transferred *in situ* into the STM head for characterization. STM topographic images were acquired in the constant-current mode using polycrystalline Pt-Ir tips. The tips were cleaned by electron-beam heating under UHV conditions and calibrated on MBE-grown Ag films prior to STM and STS measurements. Differential conductance spectra were recorded using a standard lock-in technique with a bias modulation of 50 μV at a frequency of 911 Hz.

$SrSn_3$ crystallizes in the trigonal ($R\bar{3}m$) space group and consists of 12 Sr-Sn layers per unit cell stacked along the crystallographic $c$ axis [Fig. 1(a), top]. Within each atomic layer, Sn atoms form a kagome lattice, while Sr atoms occupy the centers of the hexagons, forming a triangular lattice [Fig. 1(a), bottom]. This kagome system exhibits a superconducting transition temperature of $T_c$ = 5.5 K and non-trivial Berry phases revealed by de Haas-van Alphen (dHvA) quantum oscillations [28,29]. Together with theoretical predictions of an odd $Z_2$ invariant, these results establish $SrSn_3$ as a kagome superconductor with nontrivial electronic topology. Consistent with this picture, transport measurements have revealed signatures of distinct surface and bulk superconducting channels, with the surface superconductivity exhibiting a substantially enhanced $B_{c2}$ [29]. This robust surface superconductivity has been attributed to topological surface states, motivating a direct spectroscopic investigation of its microscopic properties. To this end, we grew epitaxial $SrSn_3$ films on Si(111) substrates by MBE and characterizing them using *in-situ* STM. In the ultrathin limit, a pronounced moiré superstructure is discernible [Fig. S2 [32]], indicating a sharp and well-defined interface between the $SrSn_3$ films and the underlying Si(111) substrates. Figure 1(b) shows a typical STM topographic image of an atomically flat $SrSn_3$ surface, resolving the Sn kagome lattice with an in-plane lattice constant of 6.9 ± 0.1 Å. This observation is consistent with previous theoretical calculations predicting that the electronic state near the Fermi level ($E_F$) is dominated by Sn $p$ orbitals [28]. No charge-density modulation is observed down to 0.4 K, distinguishing $SrSn_3$ from the extensively studied $AV_3Sb_5$ kagome superconductors [25,26,29] and making it a comparatively clean platform for investigating superconductivity without the complications of competing charge order.

Figure 1(c) shows representative d$I$/d$V$ spectra acquired on a 78-nm-thick epitaxial $SrSn_3$ film at 0.4 K. These spectra exhibit apparent superconducting gaps with pronounced coherence peaks and negligible spatial variation across the kagome lattice. Spectra measured on both the Sr and Sn sites are well described by an isotropic *s*-wave Dynes function (red lines) [35]. As the temperature increases, the superconducting gap gradually closes and becomes indistinguishable above 4.2 K [Fig. 1(d)]. The extracted gap magnitude as a function of temperature is summarized in Fig. 1(f) and follows the BCS prediction remarkably well. Fitting the temperature dependence of the superconducting energy gap, $\Delta_s(T)$, to the BCS gap function yields a zero-temperature gap $\Delta_s(0) \approx 0.68$ meV and a transition temperature $T_c \sim 4.2$ K [36]. The corresponding reduced gap ratio, $2\Delta_s(0)/k_BT_c \sim 3.8$ ($k_B$ is the Boltzmann constant), is close to the weak-coupling BCS value. Notably, the extracted $T_c$ agrees well with the surface superconducting channel [29], while remaining lower than the bulk value of approximately 5.5 K [28,29]. Together with the previously reported larger bulk superconducting gap (1.1 ~ 1.2 meV) and its strong-coupling character [28,29], these results indicate that the superconducting gap observed is predominantly associated with the surface superconducting channel, which we denote as $\Delta_s$.

Unexpectedly, the superconducting spectra evolve systematically with the tunneling junction resistance, $R_N = V/I$ [Fig. 2(a)], where $V$ and $I$ are the sample bias and tunneling current used to stabilize the STM tip prior to the *dI/dV* measurements, respectively. At large $R_N$ (> 50 MΩ), the d$I$/d$V$ spectra are characterized by the surface superconducting gap, $\Delta_s$, consistent with those shown in Fig. 1. As $R_N$ is progressively reduced, however, the spectral lineshape changes dramatically, developing a much larger fully opened superconducting gap. Figure 2(b) summarizes the evolution of the positive- and negative-bias coherence-peak energies with $R_N$, from which the gap magnitude, $\Delta^{pp}$, is estimated as half of the peak-to-peak energy separation. At sufficiently small $R_N$, $\Delta^{pp}$ reaches approximately 1.87 meV, corresponding to a superconducting gap of 1.4 ± 0.1 meV obtained from an isotropic *s*-wave Dynes fit to the top spectrum in Fig. 2(a). This value appears to be larger than $\Delta_s$ and is comparable to the bulk superconducting gap inferred from specific-heat measurements [33,34]. Combined with its exclusive appearance in the low-$R_N$ regime, this larger superconducting gap is most naturally accounted for by the reduced tip-sample distance, which preferentially enhances the tunneling probability into the more spatially extended bulk states. We therefore attribute it to the bulk superconducting state and term it as $\Delta_b$. Notably, the gap magnitude remains nearly constant in both the large-$R_N$ and small-

$R_N$ limits, while evolving continuously between the two regimes. This behavior provides direct evidence for the coexistence of two distinct superconducting channels, $\Delta_s$ and $\Delta_b$, whose relative spectral weights can be continuously tuned by $R_N$. The apparent separation between $\Delta_s$ and $\Delta_b$ [Fig. 2(c)], together with the excellent spatial homogeneity of the characteristic d$I$/d$V$ spectra over the entire $R_N$ range [Fig. S3 [32]], indicates that the observed spectral evolution is intrinsic to $SrSn_3$.

To further verify the surface and bulk origins of the two superconducting gaps, we performed systematic tunneling spectroscopy on epitaxial $SrSn_3$ films with different thicknesses. Remarkably, the two-gap behavior is universally observed across all films investigated [Fig. S4 [32]], with the principal differences manifested in $\Delta_b$ and critical tunneling junction resistance, $R_c$, that separates the two spectroscopic regimes. Figure 2(d) summarizes the film thickness dependence of $R_c$, the surface gap ($\Delta_s^{pp}$) and the bulk gap ($\Delta_b^{pp}$). To avoid uncertainties associated with Dynes fitting and enable direct comparison among different samples, we use half of the coherence peak-to-peak separation, $\Delta^{pp}$, as a measure of the superconducting gap magnitude. The surface gap, $\Delta_s^{pp}$, remains nearly unchanged over the entire thickness range, consistent with its surface origin. In contrast, $\Delta_b^{pp}$ decreases by approximately 31% as the film thickness is reduced from ~ 78 nm to ~ 5 nm, demonstrating a pronounced thickness dependence of the bulk superconducting state. These contrasting thickness dependences provide further evidence that $\Delta_s^{pp}$ and $\Delta_b^{pp}$ originate from distinct surface and bulk channels, respectively. Furthermore, $R_c$ decreases significantly with decreasing film thickness. This behavior is consistent with the progressively reduced spectral weight of the bulk superconducting channel in thinner $SrSn_3$ films. Consequently, a smaller tip-sample distance $d$ (corresponding to a lower $R_N$) is required to modify the STM tunneling matrix elements sufficiently to enhance the bulk contribution, thereby shifting the crossover between the surface- and bulk-dominated tunneling regimes to lower $R_c$.

Within the conventional BCS framework [36], the superconducting coherence length ($\xi$) scales inversely with the superconducting gap ($\Delta$), according to $\xi \sim \hbar v_F/\Delta$, where $\hbar$ and $v_F$ denote the reduced Planck constant and Fermi velocity, respectively. Consequently, a smaller $\Delta$ is generally expected to correspond to a longer $\xi$ and hence a lower upper critical field, $B_{c2}$. This simple relationship, however, does not necessarily hold when two superconducting gaps originate from distinct surface and bulk electronic channels. In particular, a surface superconducting channel may exhibit a substantially enhanced $B_{c2}$ despite its smaller gap magnitude because of its reduced dimensionality and distinct electronic structure. To test this scenario, we performed magnetic-field-dependent d$I$/d$V$ measurements and imaged vortices under magnetic fields ($B$) applied perpendicular to the sample surface. Figure 3(a) shows representative zero-bias conductance (ZBC) maps over a field of view of 240 nm × 240 nm. Bright regions with enhanced ZBC correspond to individual vortex cores, whose density increases systematically with increasing $B$. Figures 3(b,c) present the magnetic-field evolution of the *dI/dV* spectra acquired between neighboring vortices in the large-$R_N$ and small-$R_N$ limits, respectively. In the large-$R_N$ regime, where the spectra are dominated by the surface superconducting channel, the superconducting gap $\Delta_s$ gradually fills with increasing $B$ and become nearly indistinguishable above 0.7 T. Extrapolating the linear field dependence of the extracted ZBC to the normal-state value yields a surface critical field of $B_{c2,s} \approx 0.82$ T. By contrast, in the small-$R_N$ limit, where the bulk superconducting channel dominates, the larger gap $\Delta_b$ is rapidly suppressed and disappears at a substantially lower magnetic field of $B_{c2,b} \approx 0.25$ T [Fig. 3(c)]. Above this field, only the surface superconducting gap remains visible and exhibits a magnetic-field evolution nearly identical to that observed in the large-$R_N$ limit [Fig. 3(b)]. The distinct upper critical fields and magnetic-field evolutions of $\Delta_s$ and $\Delta_b$ are reproducibly observed over a broad range of $R_N$ [Figs. S5 and S6 [32]] and for all film thicknesses studied. These findings are consistent with a substantially higher upper critical field for the surface superconductivity. This provides an independent verification of the distinct surface and bulk origins of the two superconducting gaps, complementing the $R_N$- and thickness-dependent measurements.

Next, we investigate the evolution of the vortex-core states under different tunneling conditions. Figure 4(a) presents representative d$I$/d$V$ spectra acquired at the vortex-core center in the large-$R_N$ regime for $SrSn_3$ films with different thicknesses (top panel). In thicker films, a pronounced ZBCP is clearly resolved, whereas its spectral weight is progressively suppressed as the film thickness decreases. In the ultrathin limit, the ZBCP eventually evolves into a nearly featureless DOS near $E_F$. By contrast, the vortex-core spectra acquired in the small-$R_N$ regime (bottom panel of Fig. 4(a)) are characterized by a pronounced suppression of the DOS around $E_F$, reminiscent of the vortex-core spectra reported in unconventional cuprate and fulleride superconductors [37-39]. This striking contrast not only establishes the STM tunneling junction itself as a powerful tuning knob of vortex spectroscopy, but also uncovers a direct correlation between the emergence of the ZBCPs and the surface superconductivity. More importantly, the systematic suppression of the $\Delta_s$-associated ZBCP with reducing film thickness suggests a pronounced sensitivity of the vortex-bound electronic states to dimensional confinement, and possibly to the enhanced interfacial scattering of $SrSn_3$ films in the ultrathin limit.

The distinct vortex-core states associated with the surface and bulk superconducting channels are further corroborated by line-cut *dI/dV* spectra acquired across the vortex-core center ($r = 0$), as shown in Figs. 4(b) and S7 [32]. Importantly, the ZBCPs exhibit no discernible splitting upon moving away from the vortex-core center, a behavior further confirmed by the negative second derivative of the *dI/dV* spectra [Fig. 4(c)] and systematic Gaussian analysis of the ZBCPs [Fig. S8 [32]]. Both the peak position and linewidth remain essentially unchanged with increasing distance *r* from the vortex-core center, revealing the remarkable robustness of the ZBCP. Although these ZBCPs share phenomenological similarities with the zero-energy vortex-core modes reported in several candidate TSCs [13-20], their pronounced suppression with decreasing $R_N$ demonstrates an unexpected sensitivity to the tunneling conditions [Fig. 4(d)].

Several scenarios may account for the unusual vortex-core spectra observed in $SrSn_3$. One possibility is that, in the small-$R_N$ regime, the d*I*/d*V* spectra become increasingly dominated by the bulk superconducting channel, whose vortex-core response is characterized by a zero-bias suppression and can therefore reduce the relative visibility of the surface-associated ZBCPs. Alternatively, the observed behavior may be related to conventional Caroli-de Gennes-Matricon (CdGM) bound states [40], whose spectral weight has recently been predicted to be susceptible to tunneling-induced dissipation [41,42]. As $R_N$ is reduced, increased dissipation associated with the tunneling process may therefore significantly suppress the vortex-bound states. However, conventional CdGM bound states are generally expected to exhibit a clear spatial splitting upon moving away from the vortex-core center [43-45], whereas no such splitting is resolved in our measurements [Figs. 4(b,c)]. The absence of such an apparent spatial splitting disfavors some simple clean-limit CdGM scenarios, but it does not by itself exclude junction-induced dissipative renormalization effects. Whether an intrinsic splitting exists below the current experimental resolution remains an open question that calls for future measurements at lower temperatures.

Other mechanisms may also contribute to the behavior of vortex-core states. For example, the enhanced bulk contribution at low $R_N$ may increase the hybridization between the surface and bulk superconducting channels, leading to a redistribution of the low-energy d*I*/d*V* spectral weight and thereby the vortex-core states. Although this tip-induced perturbation of the superconducting state cannot be completely excluded, the nearly unchanged superconducting gap magnitudes and characteristic spectral line shapes in both the large-$R_N$ and small-$R_N$ limits argue against this possibility as the primary origin of the observed phenomena. Taken together, the present results indicate that the vortex-bound states cannot be understood independently of the coexistence of surface and bulk superconducting channels. Instead, their spectroscopic signatures are strongly influenced by the relative spectral weights of the two superconducting channels, highlighting the need for a theoretical framework that explicitly incorporates the coexistence of surface and bulk superconducting channels when interpreting the observed vortex-bound states.

In summary, we have achieved direct spectroscopic disentanglement of coexisting surface and bulk superconducting channels in the kagome superconductor $SrSn_3$. This establishes a compelling correspondence between the two superconducting channels and their distinct superconducting gaps, upper critical fields, and vortex-bound electronic states. Robust non-split ZBCPs are found to be uniquely associated with the surface superconducting channel, whereas the bulk channel exhibits pronounced zero-bias suppression inside vortex cores. These findings suggest that a non-split ZBCP alone does not constitute sufficient evidence for Majorana zero modes in multichannel superconductors. More broadly, our work establishes channel-selective tunneling spectroscopy as a general framework for separating multiple superconducting channels and exploring their interplay in topological quantum materials.

**Acknowledgments**

The work was financially supported by the National Key Research and Development Program of China (Grant No. 2022YFA1403100), the Natural Science Foundation of China (Grant No. 12474130, Grant No. 12134008, and Grant No. 52388201), and the Innovation Program for Quantum Science and Technology (Grant No. 2021ZD0302502).

**Data availability.** The data that support the findings of this study are available from the corresponding authors upon reasonable request.

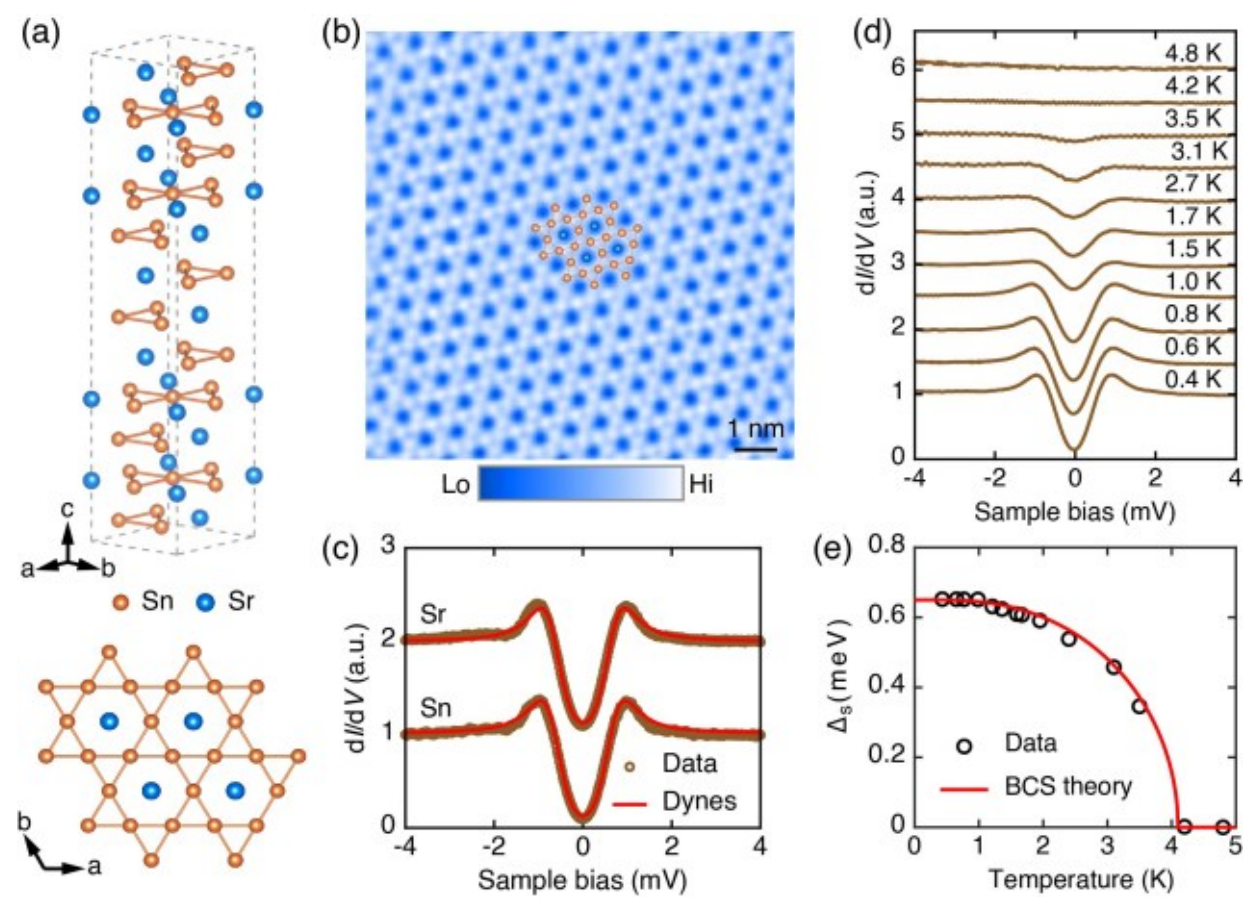

**Figure 1** (a) Crystal structure of $SrSn_3$. Upper panel: side view along the crystallographic *c* axis. Lower panel: top view of a Sr-Sn layer, showing the kagome and triangular lattices formed by Sn and Sr atoms, respectively. (b) Constant-current STM topographic image (10 nm × 10 nm, $V = 10$ mV, $I = 3$ nA) of a 78-nm-thick $SrSn_3$ epitaxial film, with orange balls indicating the Sn atomic positions. (c) Representative d*I*/d*V* spectra acquired on the Sr and Sn sites. Solid red curves are fits to an isotropic *s*-wave Dynes function. Setpoint: $V = 5$ mV, $I = 0.5$ nA. (d) Temperature-dependent d*I*/d*V* spectra measured between 0.4 and 4.8 K. Setpoint: $V = 5$ mV, $I = 0.1$ nA. (e) Temperature dependence of the superconducting gap $\Delta_s$ extracted from the Dynes analysis. The solid curve represents a fit to the BCS gap function.

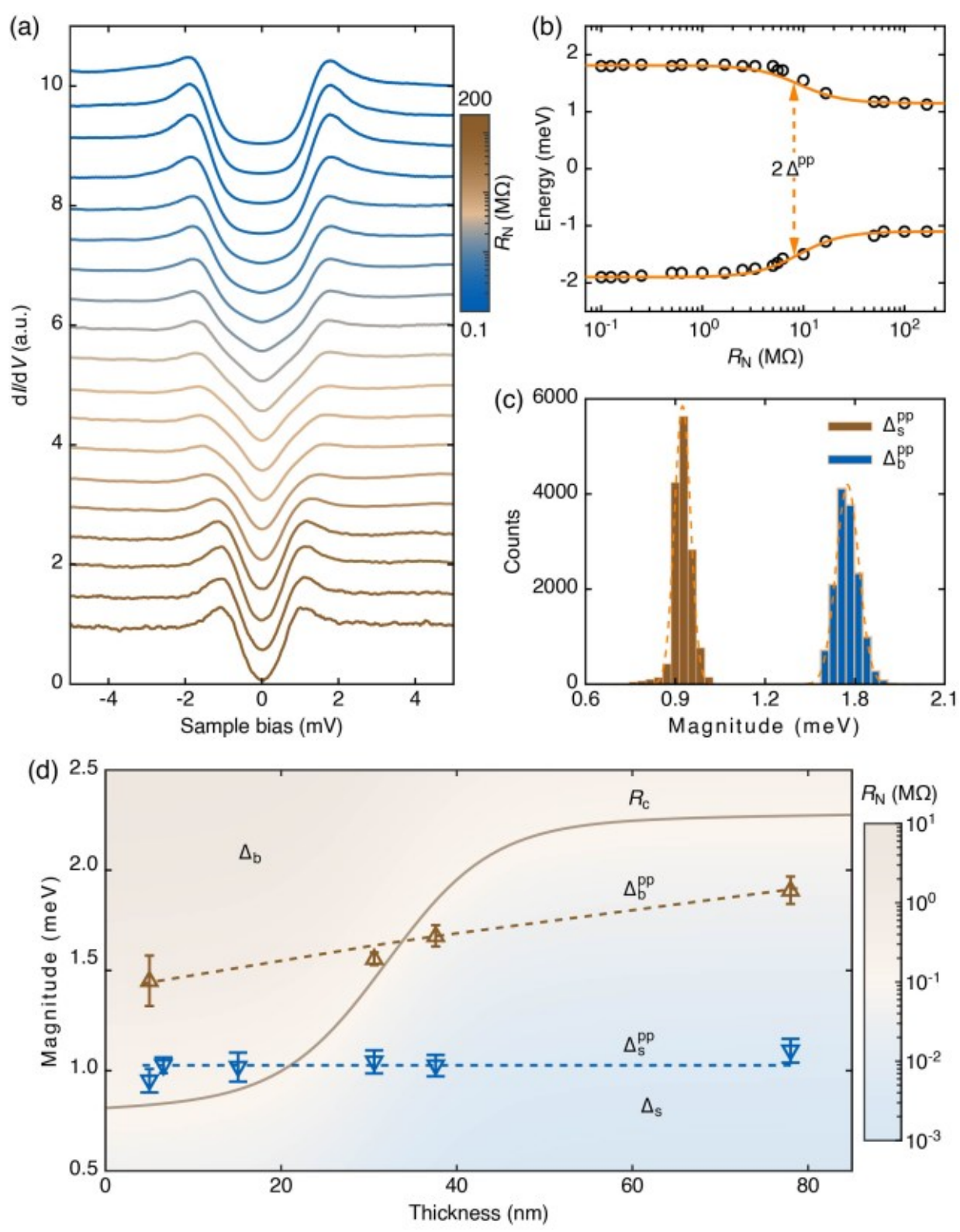

**Figure 2** (a) Evolution of the low-energy d$I$/d$V$ spectra with tunneling junction resistance, $R_N$ (denoted by the color scale), measured at 0.4 K on a 78-nm-thick $SrSn_3$ film. The tunneling junction was stabilized at a fixed sample bias of $V = 5$ mV while the tunneling current was varied. (b) Positive- and negative-bias coherence-peak energies as a function of $R_N$. The superconducting gap magnitude, $\Delta^{pp}$, is defined as half of the peak-to-peak energy separation. Solid curves are guides to the eye. (c) Statistical distributions of the surface and bulk superconducting gap magnitudes, $\Delta_s^{pp}$ and $\Delta_b^{pp}$, in the 78-nm-thick $SrSn_3$ film extracted from the large-$R_N$ and small-$R_N$ limits, respectively. Dashed curves represent Gaussian fits to the corresponding gap distributions. (d) Thickness dependence of $\Delta_s^{pp}$, $\Delta_b^{pp}$, and the characteristic tunneling junction resistance $R_c$ (marked by solid gray curve), separating the surface- and bulk-dominated tunneling regimes. Dashed lines are guides to the eye.

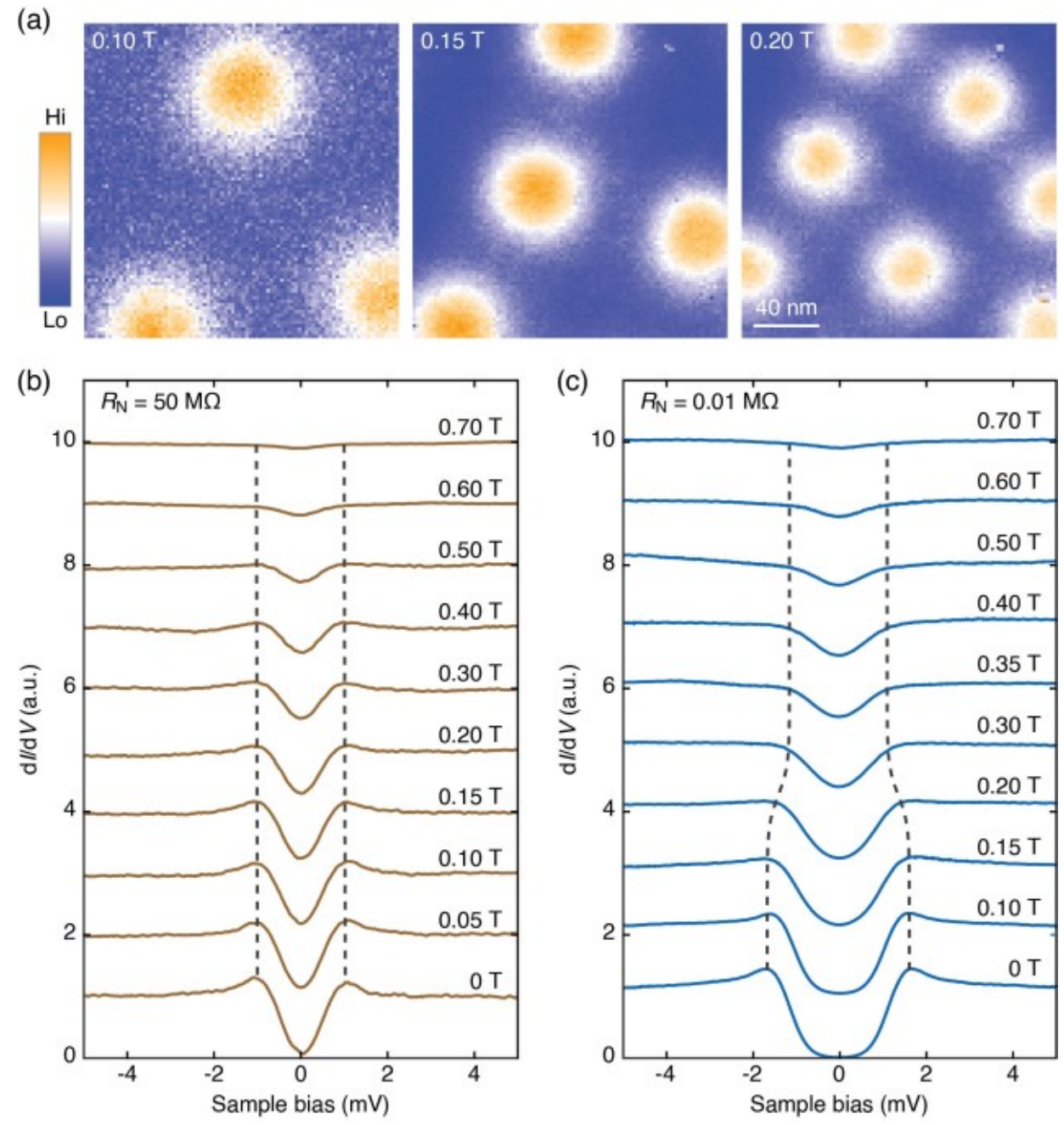

**Figure 3** (a) Zero-bias conductance (ZBC) maps acquired over a 240 nm × 240 nm field of view under out-of-plane magnetic fields of 0.10, 0.15, and 0.20 T. Bright regions correspond to individual magnetic vortex cores. Setpoint: $V$ = 10 mV, $I$ = 0.5 nA. (b,c) Magnetic-field-dependent d$I$/d$V$ spectra acquired at the same location between neighboring vortices in the large-$R_N$ (50 MΩ) and small-$R_N$ (0.01 MΩ) limits, respectively. Vertical dashed lines trace the evolution of the superconducting coherence peaks with increasing magnetic field.

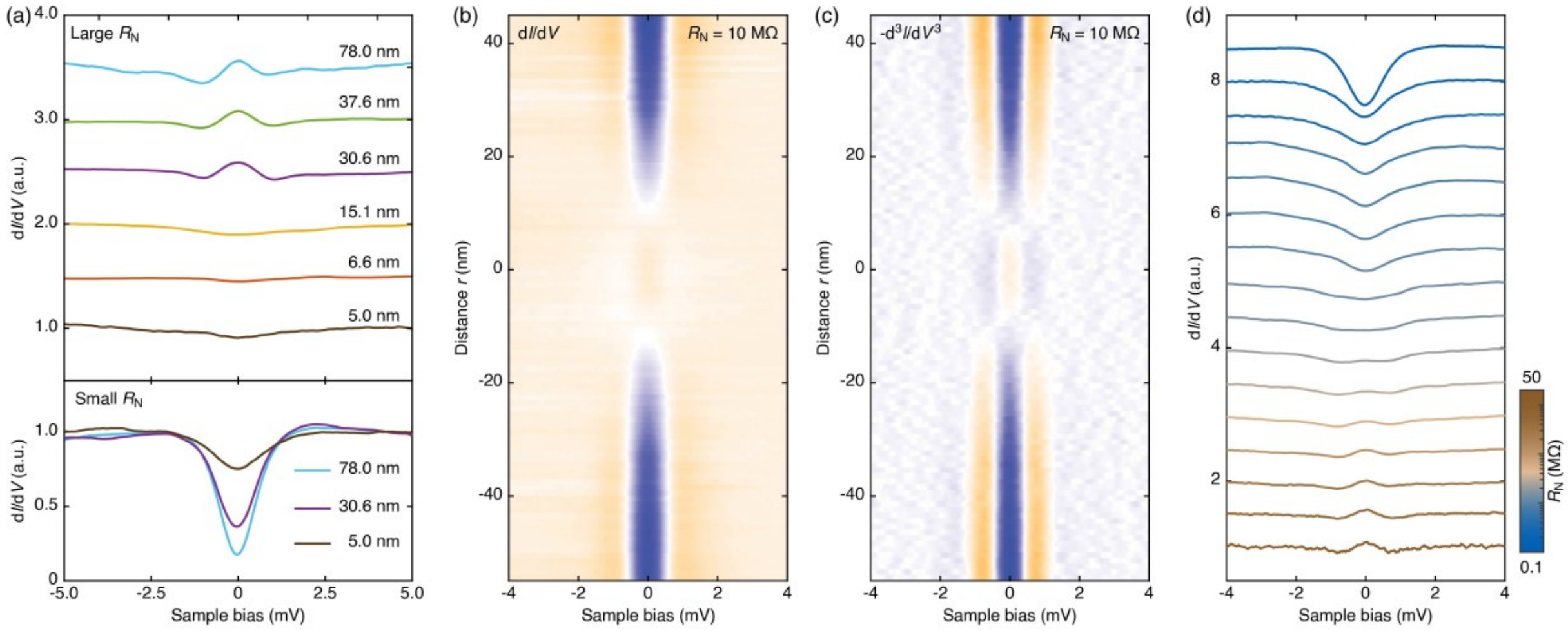

**Figure 4** (a) Representative d$I$/d$V$ spectra acquired at the vortex-core center in the large-$R_N$ regime for $SrSn_3$ films with different thicknesses (upper panel) and in the small-$R_N$ regime for selected film thicknesses (lower panel). (b,c) Intensity plots of line-cut d$I$/d$V$ spectra and their corresponding negative second derivatives (-$d^3I/dV^3$), respectively, measured along a line crossing the vortex center ($r = 0$) in the large-$R_N$ (10 MΩ) regime. (d) Evolution of the d$I$/d$V$ spectra with $R_N$, measured at the vortex-core center of the 78-nm-thick $SrSn_3$ film. The color scale represents $R_N$.

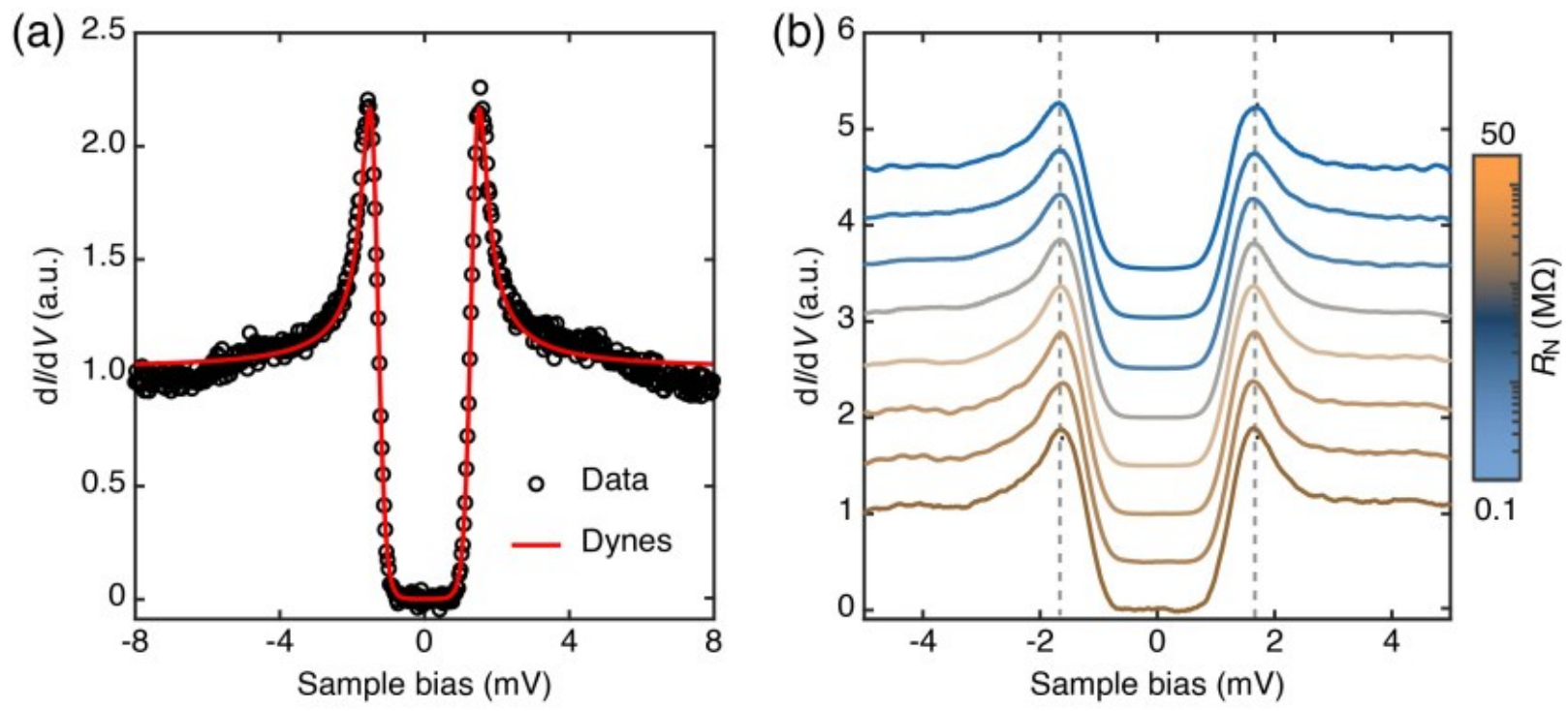

**Fig. S**1. (a) Representative low-energy d$I$/d$V$ spectrum acquired on a 50-monolayer-thick Pb film. The solid red curve is the best fit to an isotropic $s$-wave Dynes function. Setpoint: $V$ = 8 mV, $I$ = 0.5 nA. (b) Evolution of the d$I$/d$V$ spectra with tunneling junction resistance, $R_N$. No detectable change in the superconducting gap and spectral line is observed over the investigated $R_N$ range, consistent with the expected behavior of an ideal vacuum tunneling junction. Dashed lines are guides to the eye.

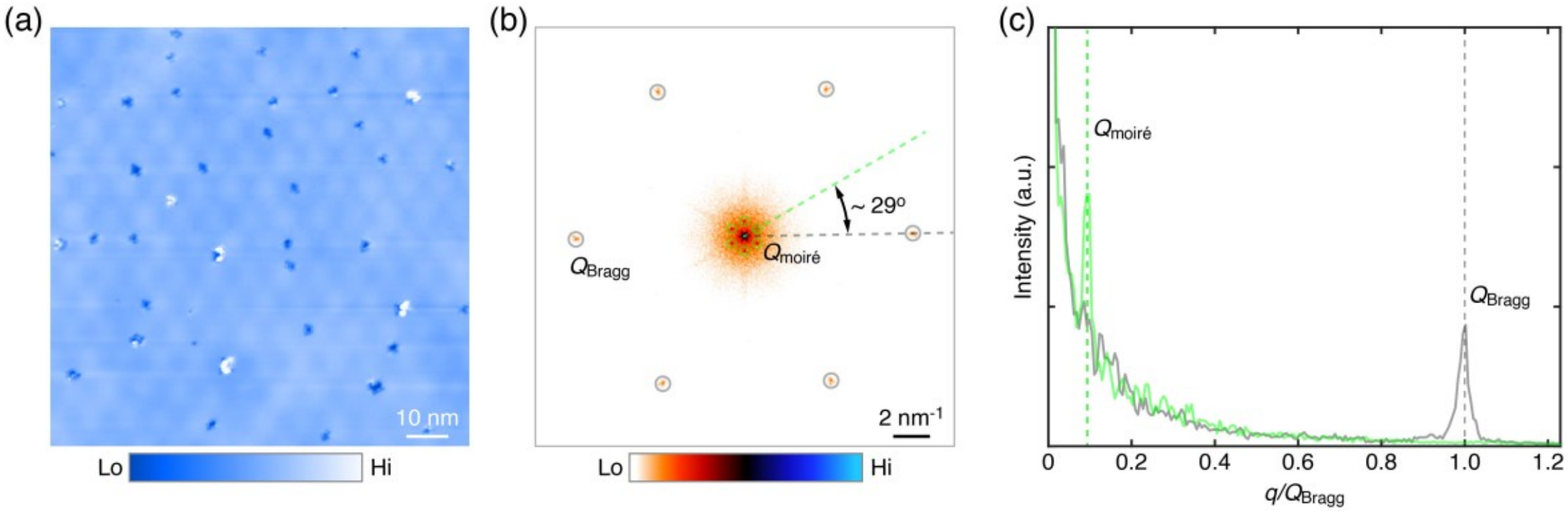

**Fig. S2.** (a) Large-area STM topographic image (100 nm × 100 nm, $V$ = -0.5 V, $I$ = 5 nA) of a 6.6-nm-thick $SrSn_3$ thin film, showing a well-defined moiré superstructure. (b) Sixfold-symmetrized fast Fourier transform (FFT) of the STM image in (a). Gray and green circles mark the Bragg peaks ($Q_{\text{Bragg}}$) of the $SrSn_3$ lattice and the moiré wave vectors ($Q_{\text{moiré}}$), respectively. The angle between $Q_{\text{moiré}}$ and a representative $Q_{\text{Bragg}}$ is measured to be approximately 30°. (c) Radial intensity profiles of the FFT along the $Q_{\text{Bragg}}$ (gray) and $Q_{\text{moiré}}$ (green) directions. The measured ratio $Q_{\text{moiré}}/Q_{\text{Bragg}} \approx 0.094$ corresponds to a real-space moiré period of 7.2 ± 0.2 nm. A simple geometric model assuming a relative lattice rotation of 13.9° between the $SrSn_3$ and Si(111) lattices yields a moiré period of ~ 7.21 nm, in excellent agreement with the experimentally observed value, and also correctly reproduces the observed orientation of the moiré wave vector.

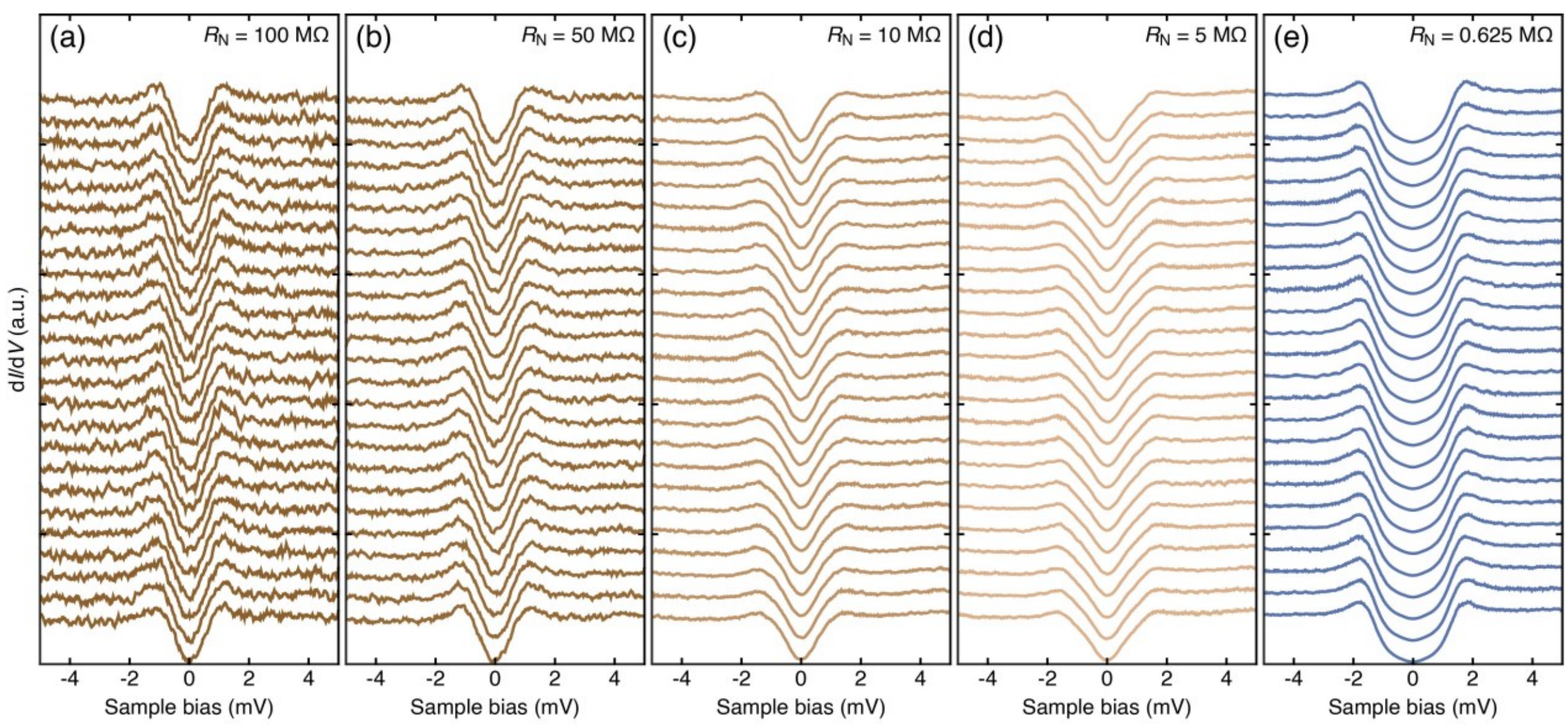

**Fig. S3.** (a-f) Spatially resolved grid d$I$/d$V$ spectra acquired over a 40 nm × 40 nm field of view at various $R_N$, as indicated. No discernible spatial variation of the low-energy d$I$/d$V$ spectra is observed at any $R_N$. Instead, the spectra evolve uniformly with $R_N$, demonstrating that the observed spectral evolution is an intrinsic and spatially homogeneous property of the $SrSn_3$ films.

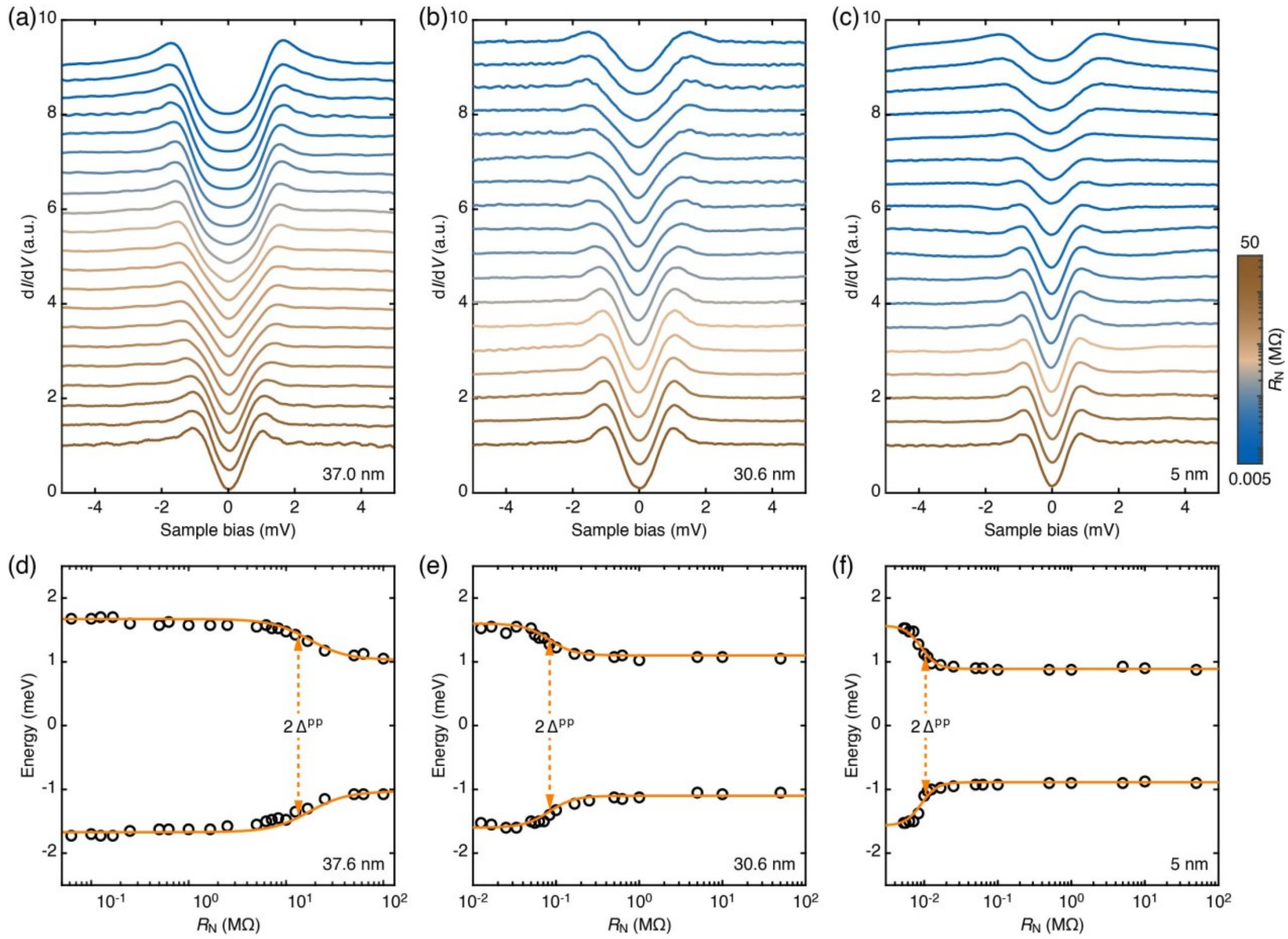

**Fig. S4.** (a-c) Evolution of the low-energy d$I$/d$V$ spectra with $R_N$, measured at 0.4 K in $SrSn_3$ thin films with thicknesses of 37.0 nm, 30.6 nm, and 5.0 nm, respectively. The color scale represents $R_N$. (d-f) Coherence-peak energies as a function of $R_N$, extracted from the spectra shown in (a-c), respectively. The superconducting gap magnitude, $\Delta^{pp}$, is defined as half of the peak-to-peak energy separation. Solid curves are guides to the eye.

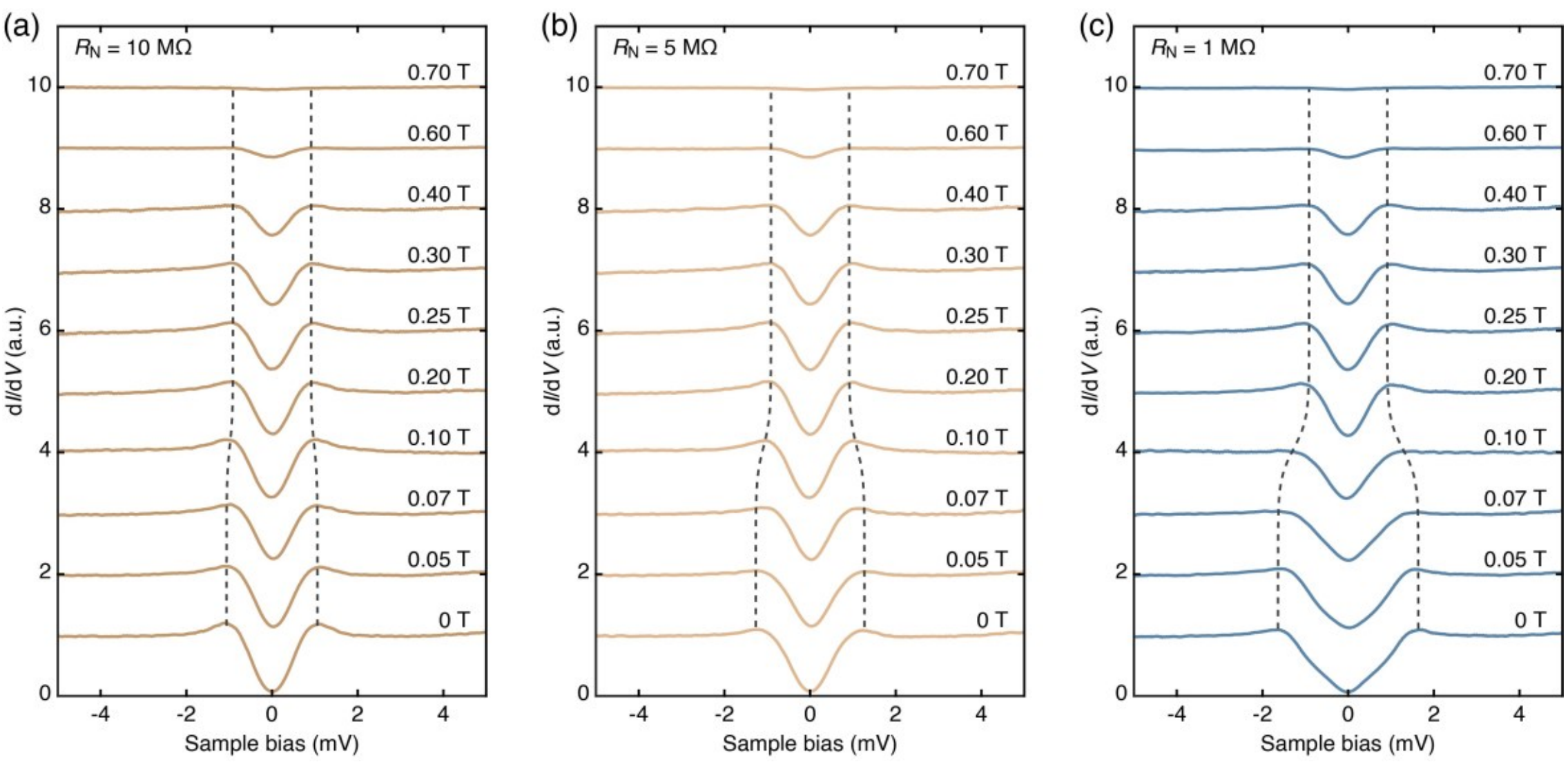

**Fig. S5.** (a-c) Magnetic-field-dependent d$I$/d$V$ spectra acquired at the same location between neighboring vortices at different $R_N$, as indicated. Dashed lines trace the evolution of the superconducting coherence peaks with magnetic field. The crossover from the bulk-dominated superconducting gap $\Delta_b$ to the surface-dominated gap $\Delta_s$ become more discernible at small $R_N$.

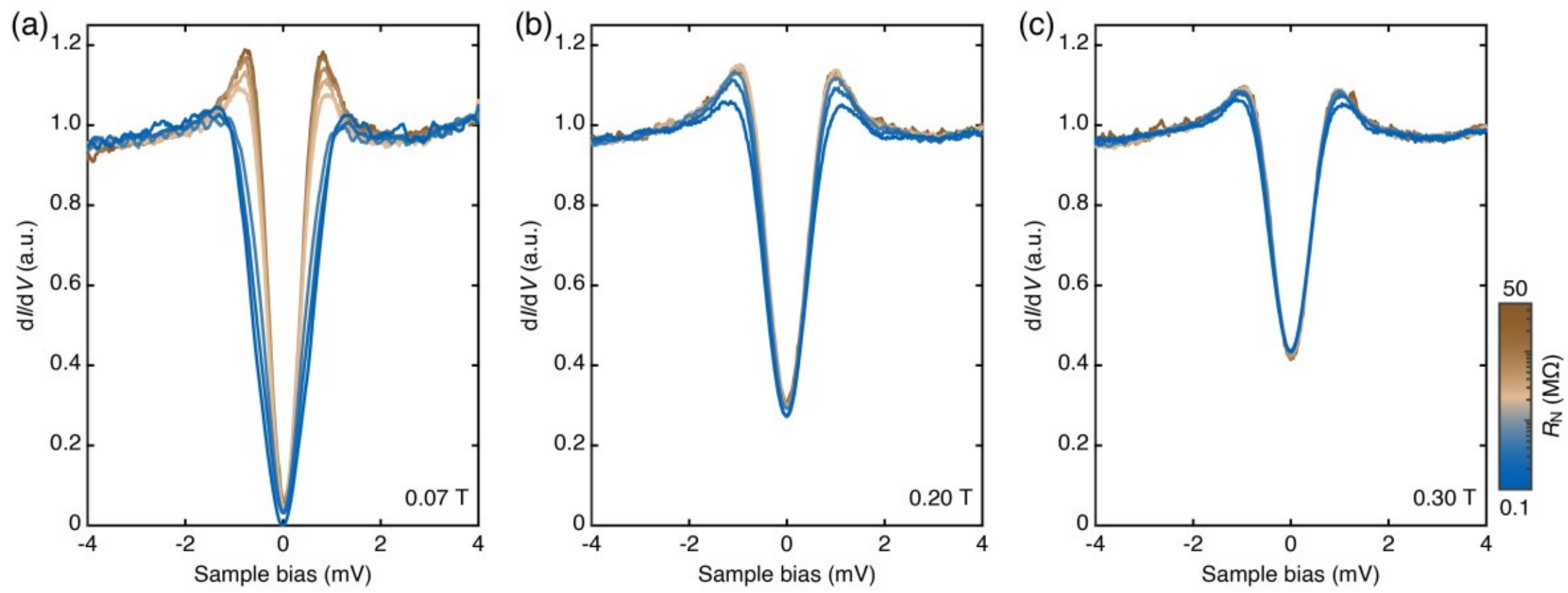

**Fig. S6.** (a) Evolution of the d$I$/d$V$ spectra with $R_N$, measured at the same location between neighboring vortices under an out-of-plane magnetic field of 0.07 T. As $R_N$ decreases, the spectra exhibit a crossover from the surface-dominated superconducting gap $\Delta_s$ to the bulk-dominated gap $\Delta_b$, similar to that observed at zero field. (b,c) Same as (a), measured under out-of-plane magnetic fields of 0.20 and 0.30 T, respectively. As the magnetic field increases, the bulk superconducting contribution is progressively suppressed, weakening the $R_N$-dependent crossover (b) and eventually eliminating it (c). The resulting d$I$/d$V$ spectra become nearly independent of $R_N$, indicating the complete suppression of the bulk superconducting channel.

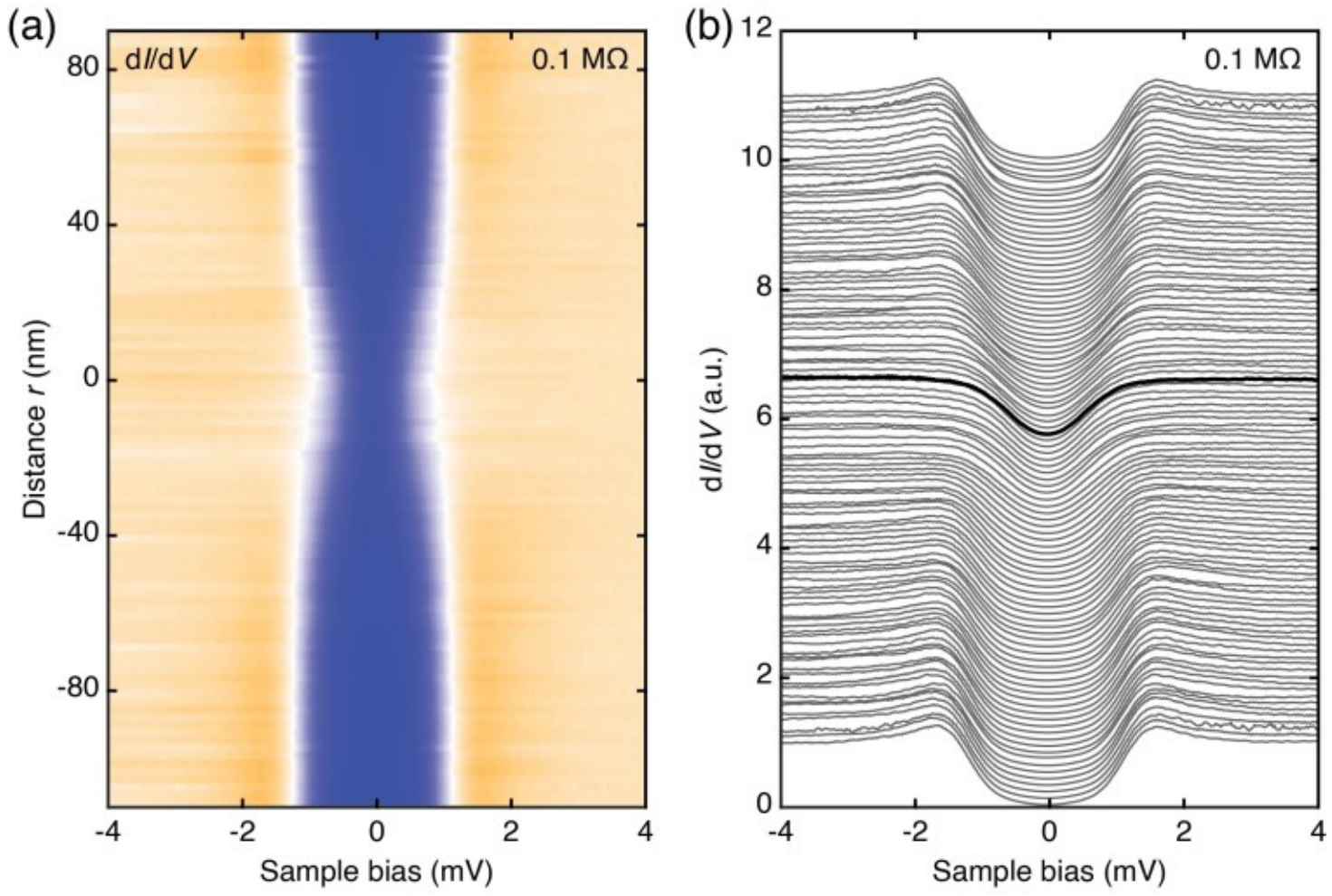

**Fig. S7.** (a) Intensity plot of the line-cut d$I$/d$V$ spectra measured along a 200 nm trajectory crossing the vortex center ($r = 0$) in the small-$R_N$ limit ($R_N = 0.1$ MΩ). (b) Waterfall plot of the spectra in (a), with the black curve showing the d$I$/d$V$ spectrum at the vortex center. Unlike the large-$R_N$ regime, where a pronounced ZBCP is observed, the vortex-core spectrum in the small-$R_N$ limit exhibits a pronounced zero-bias suppression.

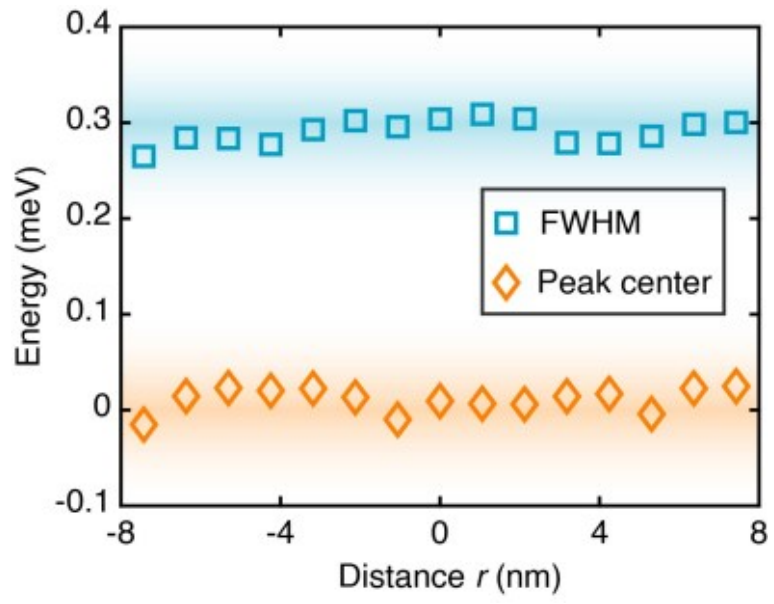

**Fig. S8.** Spatial dependence of the peak energy (diamonds) and full width at half maximum (FWHM, squares) extracted from single-Gaussian fits to the ZBCPs measured across an individual magnetic vortex. The peak energy remains pinned near zero within experimental uncertainty, while the FWHM shows no discernible variation over the measured spatial range.